\documentclass[english,a4paper,12pt]{article}
\usepackage{babel}
\usepackage[dvips]{graphicx}
\usepackage[latin1]{inputenc}
\usepackage{natbib}
\usepackage{amsmath,amssymb,
}

\newtheorem{theorem}{Theorem}
\newtheorem{definition}{Definition}

\newtheorem{remark}{Remark}

\begin{document}

\title{\large High dimensionality: The latest challenge to data analysis}

\author{ \small
A. M. Pires and J. A. Branco \\
\small
Department of Mathematics and CEMAT, IST, 
Universidade de Lisboa, \\ 
\small 1049-001 Lisboa, Portugal \\
\small \{apires, jbranco\}@math.ist.utl.pt}
\date{}

\maketitle

\begin{abstract}
\noindent
The advent of modern technology, permitting the measurement of thousands of characteristics simultaneously, has given rise to floods of data characterized by many large or even huge datasets. This new paradigm presents extraordinary challenges to data analysis and the question arises: how can conventional data analysis methods, devised for moderate or small datasets, cope with the complexities of modern data? The case of high dimensional data is particularly revealing of some of the drawbacks.
We look at the case where the number of characteristics measured in an object is at least the number of observed objects and conclude that this configuration leads to geometrical and mathematical oddities and is an insurmountable barrier for the direct application of traditional methodologies. If scientists are going to ignore fundamental mathematical results arrived at in this paper and blindly use software to analyze data, the results of their analyses may not be trustful, and the findings of their experiments may never be validated. That is why new methods together with the wise use of traditional approaches are essential to progress safely through the present reality.
\end{abstract}

\textbf{keywords:} Curse of dimensionality; High dimensional data; Mahalanobis distance.

\section{Introduction}

When $n$ ``objects'' (patients, subjects, cells, samples, etc) are measured on $p$  distinct, possibly correlated, characteristics (or variables) we say we have a multivariate $p$-dimensional dataset. Examples can be found in all areas of science as well as in many current human activities. Statistical methods to deal with this kind of datasets were developed along the twentieth century under the assumption that the number of variables is much smaller than the number of observations    \citep[see e.g.,][]{Johnson:Wichern:2007}. However, the automatic acquisition of data due to the extraordinary development of new technologies  observed in the last decades, has changed this paradigm. 
 Nowadays it is quite common to have datasets with a number of variables much larger than the number of observations. A variety of examples of such high dimensional datasets can be found in, for instance, the genomics, chemometrics, astronomy, climate or finance literature.  How are we dealing with this new scenario?  
At least two  references \citep{Clarkeetal:2008,Johnstone:Titterington:2009}  acknowledge the difficulties encountered  and basically recognize our ignorance about the basic properties of high dimensional data spaces.
Despite this ignorance a continuous stream of new methods prepared to deal with this kind of data has flooded the scientific literature. For example, chapter 18 of \citet{Hastie:Tibshirani:Friedman:2009}  is dedicated to high dimensional problems ($p$ much larger than $n$) and describes many of the methods proposed up to 2008.

In this paper we present some new mathematical results about the geometry of datasets in high dimension which reveal interesting fatal features that have been ignored so far. 
The consequences of these findings are determinant for the correct
approach to analyze data. As a matter of fact it will become clear
that it is nonsense
to freely use many of the traditional data analysis methods directly to study high
dimensional data, in particular if $p \ge n$. The paper is organized as follows. 
In the next  section we establish the background, describe the problem and
present the main findings. The consequences of the results and possible solutions are discussed in
the concluding section.

\section{Geometric aspects of multivariate data}

\subsection{When p is small} 
In this case $n$ is usually much larger than $p$ ($p \ll n$).  If, for instance, $p=2$ the data can be represented by $n$ points in a bivariate scatter plot, where the two axis represent the two variables. The main features of the data (relationship/correlation between the variables, grouping of objects, outliers, and others) can be visualized on this scatter plot. When $p=3$ things are not so easy. We may produce a 3-d scatter plot, but this scatter plot is ultimately represented in a two dimensional space (computer screen/sheet of paper) and what we really observe is a projection of the 3-d point cloud onto a 2-d subspace. This can be informally described as a picture of the point cloud taken from a given position and angle. To understand a 3-d dataset we must take many pictures from varying positions and angles, which means that 
visualization of the data is, no doubt, more difficult when $p=3$ than when $p=2$. What about when $p > 3$? Conceptually we could always apply a similar procedure, that is, project the p-dimensional data points onto 2-d subspaces.  It is easy to understand that, as the number of dimensions increases, we would need an increasingly larger number of pictures to get just a glimpse of the data. This is yet a manifestation of the ``curse of dimensionality'' \citet{Bellman:1957}. 

At this point we must recognize the need and importance of multivariate statistical methods. These  methods are designed to analyze datasets with $n$ observations on $p$ variables organized in a data matrix, $X=\{x_{ij}\}_{i=1, \ldots , n; j=1, \ldots , p}$, with $n$ rows and $p$ columns, where $x_{ij}$ denotes the value of the $j$th variable for the $i$th object.  We restrict our attention to numerical variables and we assume throughout that the $n$ points are in ``general position'', which means that there are no redundancies in the data, like, for instance, two exact replicas of an observation (a precise definition is given in  Appendix A).

 Multivariate statistical methods (considered in a broad sense, i.e., including machine learning and related topics) can extract and quantify relevant features of the data, and also produce, in many cases, two dimensional graphical representations to complement the analysis. Most of those methods use matrix algebra techniques. For instance, principal components \citep{Hotelling:1933}, which are successive uncorrelated directions with maximal variance. Principal components are defined by the eigenvectors of the covariance matrix, $S$,  or of the correlation matrix, $R$, whereas the variances along the principal components  can be obtained by computing the corresponding eigenvalues. A very important tool in multivariate analysis, which is related to a number of methods, including principal components, is the Mahalanobis distance \citep{Mahalanobis:1936}. 
 
When comparing/analyzing data on a single variable it is often informative to compute the distance of each observation to the center of the data (usually identified by the sample mean) or the distance between pairs of observations, taking into consideration the intrinsic variability of the data, usually measured by the standard deviation. Recall that for a dataset consisting of $n$ observations on one variable, $(x_1, \ldots , x_n)$, the $z$-scores are $z_i=(x_i - \bar{x})/s$, where $\bar{x}$ and $s$ are the arithmetic mean and standard deviation of the $n$ observations. The $z$-scores have been used for instance to detect outliers, though they have a number of important limitations \citep{Barnett:Lewis:1994}. The mean  of $(z_1, \ldots , z_n)$ is zero while both its standard deviation and variance are equal to 1. Also, $|z_i|=\{(x_i - \bar{x})^2/s^2 \}^{1/2}$ is both the standardized distance between $x_i$ and $\bar{x}$ and the Euclidean distance between $z_i$ and $\bar{z}\equiv 0$, whereas the Euclidean distance between $z_i$ and $z_j$, $|z_i-z_j|$, is equal to $\{(x_i - x_j)^2/s^2\}^{1/2}$, the standardized distance between $x_i$ and $x_j$. 

Consider now a multivariate $n \times p$ dataset and imagine a standardizing transformation with similar properties as the $z$-scores transformation. Computing the $z$-scores separately for each variable does not achieve that desideratum unless all the variables are uncorrelated (that is, in case both $S$ and $R$ are diagonal matrices).
It turns out that the multivariate transformation which standardizes a multivariate dataset taking into account the correlations between all the variables is the Mahalanobis transformation, defined in matrix notation by ${z}_i=S^{-1/2}({x}_i - \bar{x})$, where ${x}_{i}$ is a vector containing the $p$ measurements of the $i$th object, $\bar{x}$ is the mean vector, containing the $p$ means of the individual variables and $S^{-1/2}$ is a square root of the inverse of $S$. This transformation has the same form of the $z$-scores transformation, but uses matrices in place of single values. 
It is easy to verify that the standardized data matrix, $Z$, formed with the ${z}_i$ vectors, is such that all the variables have zero mean and unit variance, and all pairwise covariances are zero (in other words, the mean vector of $Z$ is the null vector,  and both its covariance and correlation matrices are the identity matrix, $I$). The Euclidean distance between a standardized observation, ${z}_i$, and the null vector is $\left( \sum_{j=1}^p  z_{ij}^2 \right)^{1/2}$, which can be written in matrix notation as
$\left( {z}_i^T{z}_i \right)^{1/2} = \left\{ \left({x}_i - \bar{x} \right)^T S^{-1} \left({x}_i - \bar{x} \right) \right\}^{1/2}$ and defines the Mahalanobis distance between the observation ${x}_i$ and the mean of the $n$ observations,  $d_{{x}_i,\bar{x}}$. The Euclidean distance between two standardized observations, ${z}_i$ and ${z}_j$, is $\left\{ \left({z}_i- {z}_j\right)^T\left({z}_i-{z}_j\right) \right\}^{1/2}= \left\{ \left({x}_i - {x}_j\right)^T S^{-1}\left({x}_i - {x}_j\right)\right\}^{1/2}$ and defines the Mahalanobis distance between ${x}_i$ and ${x}_j$, $d_{{x}_i,{x}_j}$. Like the $z$-scores, Mahalanobis distances to the mean have been used to detect outliers in multivariate datasets \citep{Gnanadesikan:Kettenring:1972}. Mahalanobis distances between two objects are useful, {\it e.g.}, in clustering applications. 
 
\subsection{When p is large and n is small} 

Before we discuss the very complicated issue of the visualization of this kind of data, let us consider an apparently naive question: can we use Mahalanobis distance when the number of variables is larger than the number of observations? 
A first quick answer would be: no, because when $p \ge n$ the covariance matrix can not be inverted ($S$ is singular) and therefore Mahalanobis distance is not defined.  However, it is still possible to define the Mahalanobis distance by reasoning as follows. Three points (in general position) in a 3-d space ($p=3$, $n=3$) define a plane, in other words, they lie on a certain 2-d subspace. If we adopt a coordinate system in that subspace, Mahalanobis distances can be computed, because the new covariance matrix will not be singular. A similar argument can be applied to higher dimensional spaces.
A dataset with $n$ observations in $p$ variables, with $p \ge n$, can be represented, without any loss of information, in a new set of $q=n-1$ variables, for which Mahalanobis distances can be computed. Moreover, such a set of variables is easy to find. Consider, for instance, the $n-1$ principal components of $X$ corresponding to non-null eigenvalues of $S$. These new variables are linear combinations of the original variables and it is easy to move from one system to the other. 
The Mahalanobis distance defined in this way coincides with the generalized Mahalanobis distance \citep{Mardia:1977}, for which the non-existing inverse is replaced by the Moore--Penrose generalized inverse.
Therefore we conclude that we can still standardize datasets with $p \ge n$, by applying the Mahalanobis transformation in the largest subspace for which the covariance matrix is invertible, that is, using $q=n-1$ new variables  which are linear combinations of the original $p$ variables. But when we perform the computations just described we arrive at the following surprising result (details and proofs  are given in Appendix A).

\begin{theorem}
For every dataset with $n$ points,  $x_1, \ldots, x_n$,  and $p \ge n-1$  variables we have that: (i)~$d_{{x}_i,\bar{x}}=(n-1)\; n^{-1/2}$, for every $i=1, \ldots , n$, and (ii)~$d_{{x}_i,{x}_j}= \left\{2(n-1)\right\}^{1/2}$ for every $i\ne j=1, \ldots , n$. In other words, whatever the points  $x_1, \ldots, x_n$, as long as $p \ge n-1$, the standardized data always form a regular pattern in which the distance from every point to the center is a constant and the distance between any two points is another constant, both constants depending only on~$n$. $\Box$
\end{theorem}

Despite Mahalanobis distance has been known and used for more than 70 years, these apparently simple results were not found in the literature.
Incidentally,  we are also able to show, following almost the same proof as in Theorem 1, that for every dataset, irrespective of the number of observations or variables, we have that (iii)~$d_{{x}_i,\bar{x}}\le (n-1)\; n^{-1/2}$, for every $i=1, \ldots , n$, and (iv)~$d_{{x}_i,{x}_j} \le \left\{2(n-1)\right\}^{1/2}$ for every $i\ne j=1, \ldots , n$ (a proof of (iii) was published in  \citet{Trenkler:Puntanen:2005} and also in \citet{Gath:Hayes:2006}.
The results given in Theorem 1 can also be connected to asymptotic results in  \citet{Ahnetal:2007}.
  
Figure~1 illustrates Theorem 1 for datasets with 3 points and a number of variables larger or equal than 2. This is the only case we can represent directly in a 2-d plot (the plane containing the three points), and we can see that every dataset with 3 points in $p \ge 2$ variables corresponds, when standardized and apart from an arbitrary rotation, to the ``same'' equilateral triangle (the general position assumption excludes cases where the 3 points lie on a straight line). Similarly, for $n=4$ points  in $p \ge 3$ variables, every dataset  corresponds, when standardized, to the ``same'' regular tetrahedron (again apart from an arbitrary rotation). For $n\ge5$ we have to imagine a regular geometric object in a $p\ge n-1 \ge 4$ dimensional space which is the appropriate member in the sequence: equilateral triangle, regular tetrahedron, \ldots These objects are called regular simplices (triangle$\,\equiv\,$2-simplex; tetrahedron$\,\equiv\,$3-simplex; pentachoron$\,\equiv\,$4-simplex; \ldots; $(n-1)$-simplex; \ldots). By the above result, we can envisage the standardized version of an $n \times p$ data matrix, with $p \ge n-1$, as the $n$ vertices of a regular $(n-1)$-simplex, such that the length of every edge is $d_{{x}_i,{x}_j}= \left\{2(n-1)\right\}^{1/2}$, and every vertex  is located at a distance of $d_{{x}_i,\bar{x}}=(n-1)\; n^{-1/2}$  from the center of the simplex. This is a very simple regular structure with all the points at the boundary of its convex hull  (the simplex) while the interior of the convex hull is completely empty.  
This last property is also shared by the original data, because they can be seen as a non-singular linear transformation of the standardized data: ${x}_i=S^{1/2}{z}_i + \bar{x}$.

\begin{figure}[htbp]
\begin{center}
\includegraphics{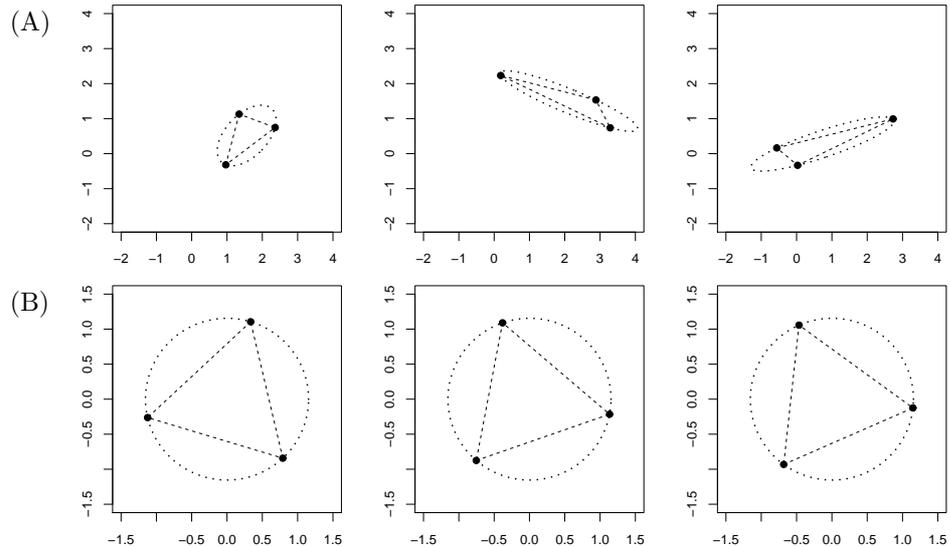}
\end{center}
\caption{Three datasets with 3 random points in dimension 2. (A) Original observations, $x_{1}$, $x_{2}$ and $x_{3}$. (B) Standardized observations, $z_i=S^{-1/2}(x_i - \bar{x})$, $i=1,2,3$.}
\label{fig1}
\end{figure}

We have thus shown that the point cloud of a multivariate dataset with $p \ge n-1$ is like an empty shell which, when standardized, looks the same whatever the data. This reality has strong implications. If we insist on considering the multivariate relation between all the variables, it is not possible, unless external information is provided,  to: (i)  separate outliers from non outlier observations, (ii) detect any kind of deviations from symmetric structures,  (iii)  distinguish between linear and non linear structures, or (iv) identify any type of clustering. 

Let us now consider  projections of these data structures, that is, imagine traveling ``around'' the point cloud and taking a large number of interesting pictures. The next result shows that these pictures can show virtually anything we want to see and have therefore to be used very carefully when extracting conclusions about the data.

\begin{theorem} For every matrix with $n$ points and $p \ge n-1$ variables, $X$, and every non-singular two dimensional arrangement of $n$ points ($Y$) it is always possible to find (explicitly) an orthogonal projection of  $X$ which is ``similar'' to $Y$.
\label{th2}
\end{theorem}

The result given in this theorem  
can be connected to some results in the mathematics literature  \citep{Baryshnikov:Vitale:1994,Eastwood:Penrose:2000}. Remarkably, the ``piling effect'', characterized by the existence of directions such that projections of data onto those directions  take  only two distinct values \citep{Ahn:Marron:2010}, is just a special case of Theorem~\ref{th2}.
 
Figure~2  illustrates Theorem~2 using two well known datasets: the colon cancer microarray data \citep{Alonal:1999} and the ORL face database \citep{Samaria:Harter:1994}. The implications for data analysis are the same as for Theorem 1: if there is no external information, there is little we can conclude about the multivariate structure of an high dimensional dataset. For instance, is a projection with an outlier evidence of the existence of an outlier? It can not be, because we can see such a projection in every high dimensional dataset with just the same number of points, irrespective of the presence of any outlying observation.

\begin{figure}[htbp]
\begin{center}
\includegraphics{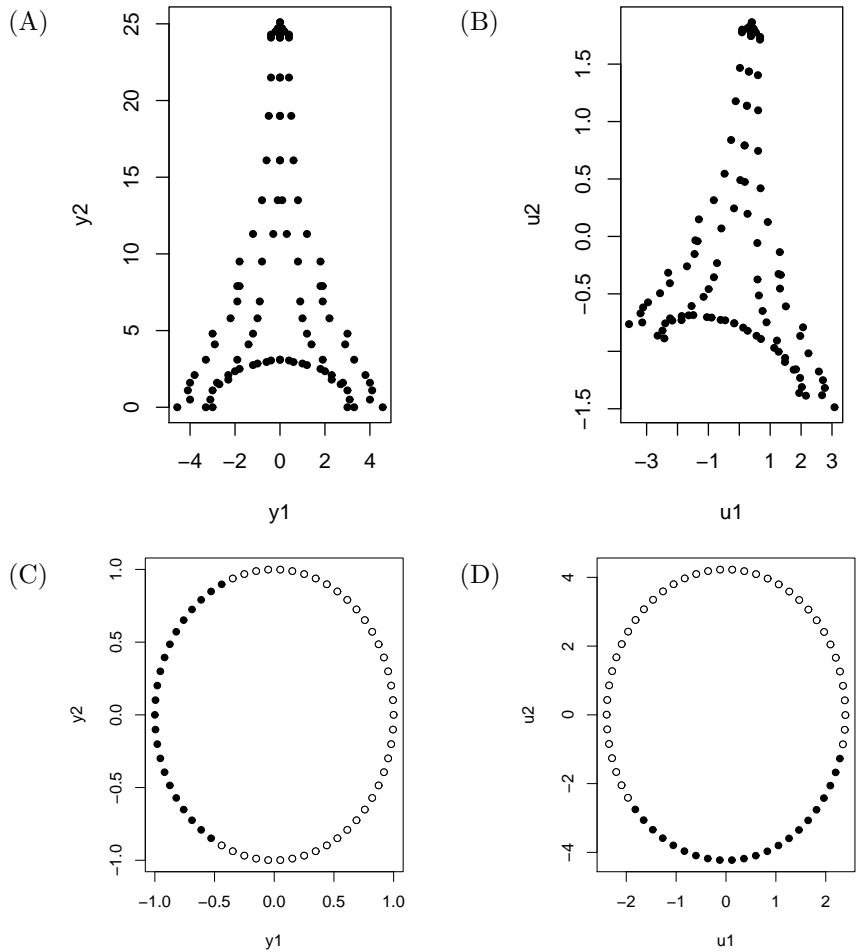}
\end{center}
\caption{Illustration of Theorem 2. (A) and (C)  Arbitrary configurations of points in 2 dimensions. (B) An orthogonal projection of high dimensional data which is similar to the configuration in (A), where the data comes from the ORL face database \citep{Samaria:Harter:1994}  whith  $n=400$ observations (pictures) on $p=10304$ variables (pixels). (D) An orthogonal projection of high dimensional data which is similar to the configuration in (C), where the data comes from the colon cancer microarray dataset \citep{Alonal:1999} whith  $n=62$ observations (arrays) on $p=2000$ variables (genes); the black circles correspond to the 22 normal tissue samples and the white circles correspond to the 40 tumor tissue samples (the arbitrary configuration in (C) specifies separation of the two groups).}
\label{fig2}
\end{figure}

We now know that the $p\ge n-1$ data space is quite odd. It is then natural to ask whether those singularities appear suddenly when $p$ reaches $n-1$ or whether the properties of the space change gradually as $p$ approaches $n-1$ from bellow. The upper bound of the Mahalanobis distance, together with the fact that the expected value of $d^2_{x_i,\overline{x}}$ is  $(n-1)p/n$ for $p \le n-1$, whatever the distribution of the data \citep{Mardia:1977},  show that a transition must start far before $p$ reaches $n-1$ \citep[this may also be related to Corollary 1.1 in][which states a rigorous result about such  transitions for samples from a multivariate normal distribution]{Donoho:Tanner:2005}.  Figure~3 illustrates this aspect. The plots show orthogonal projections of the colon cancer microarray dataset, selected to be as similar as possible to configuration (C) of Figure~2, but using only a subset with $p < n$ randomly chosen variables. Contrary to what happened in Figure~2, it is no longer possible to replicate any configuration, however, certain aspects of the given structure remain visible, even in the case $p=10$, where the separation of the two groups is still noticeable. This example shows that it may be dangerous to interpret motifs seen in projections if the ratio $n/p$ is not large enough.

\begin{figure}[t]
\begin{center}
\includegraphics{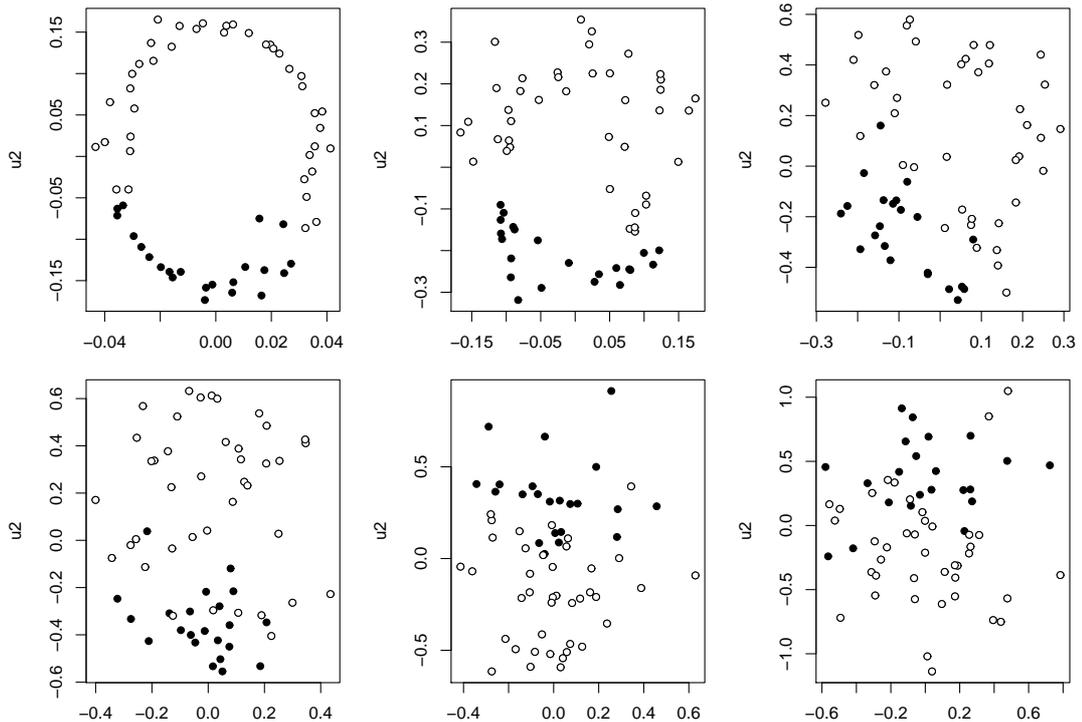}
\end{center}
\caption{Six orthogonal projections of the colon cancer microarray dataset, with \textit{p} $<$ \textit{n}, selected to be as similar as possible to the configuration (C) of Figure~\ref{fig2}. From top to bottom, left to right: \textit{p} = 60, 50, 40, 30, 20 and 10 randomly selected variables.}
\label{fig3}
\end{figure}

Taking into account  the results  presented one should investigate many  recent scientific discoveries in various areas which relied on the analysis of high dimensional datasets. This can be easily done by (i) generating a number of artificial datasets similar to the dataset under study, for instance, with the same number of variables and observations, or with equal means and variances, (ii) studying those artificial datasets in the same manner as the original data, (iii) comparing the results. One example of the application of this strategy is presented next, but others could be given. 

\subsection{Example} 
In this example we consider again the colon cancer microarray dataset and assess the performance of several outlier detecting methods recently proposed and advertised as appropriate for high dimensional datasets  \citep[there is an extensive literature about the outliers of this dataset, see][and the references therein]{Shieh:Hung:2009}. The  
methods considered are the ROBPCA of \citet{Hubert:Rousseeuw:VandenBranden:2005}, three methods proposed in \citet{Filzmoser:Maronna:Werner:2008}, denominated by the authors as PCOUT, SIGN1 and SIGN2, and the method proposed in \citet{Shieh:Hung:2009}, referred to as SH. More details about this example are provided as Supplementary Material. 

We first concluded that all the methods point out some observations as outliers (ROBPCA: 5 normal and 10 tumor; PCOUT: 1 normal and 6 tumor; SIGN1: 7 normal and 21 tumor; SIGN2: 4 normal and 8 tumor; SH: 3 normal and 5 tumor), unfortunately they are unanimous only about two cases (one normal and one tumor).

In order to show that the flagged outliers may well be illusions created by the high dimensionality, we  simulated 500 datasets with multivariate normal distribution but similar to the original colon cancer dataset  (that is, with two groups, same number of variables and observations, same means and covariances per group). We then applied the five outlier detecting methods to each dataset and registered the number of outliers detected. The median number of detections observed  were the following: in the group of 22 observations, ROBPCA: 6, PCOUT: 3, SIGN1: 8, SIGN2: 1, SH: 2, and in the group of 40 observations, ROBPCA: 12, PCOUT: 4, SIGN1: 17, SIGN2: 2, SH: 2. These figures are far too large than would be anticipated and are similar to the number of outliers ``detected'' in the real data. We must analyze the results deeper.

All the methods are calibrated at the normal, meaning that the probability of declaring an observation from a normal distribution as an outlier, is pre-specified at a certain level, $\alpha$. In this case $\alpha = 2.5\%$ for all the methods and both the real and simulated data. According to this definition, the number of outliers detected in a sample of $n$
observations from a normal distribution follows a binomial$(n, \alpha)$ distribution. Therefore, we expect that the sample containing the number of outliers found in each of the simulated datasets to behave as a sample with 500 observations from a binomial(40,0.025) or a binomial(22,0.025). From now on we refer only to the part of the simulation with $n=40$, as the results and conclusions from the other part ($n=22$) were similar.
 
The plots in the left column of Figure 4 show the observed frequency distributions of the number of outliers detected with $n=40$
and all the variables ($p=2000$) included. The '+' symbol in the plots indicate the expected frequencies under the  binomial(40,0.025) distribution. All the observed frequencies are very different from the expected frequencies. We may conclude that the detection methods are not working as they should, especially the first three, ROBPCA, PCOUT and SIGN1, and are probably providing misleading results also for the real data.

We are convinced that the responsibility for the failure of the outlier detection methods can be ascribed, at least partially, to the high dimensionality of the data. To support this claim we repeated the simulation with normal data for a much smaller number of variables. The illustration described in Figure 3 shows that $p=10$ may already be too large when $n=40$. Then we have selected randomly $p=5$ variables from the original 2000. The plots with the number of outliers detected are shown on the right column in Figure 4.

\begin{figure}[htbp]
\begin{center}
\includegraphics[height=0.9\textheight]{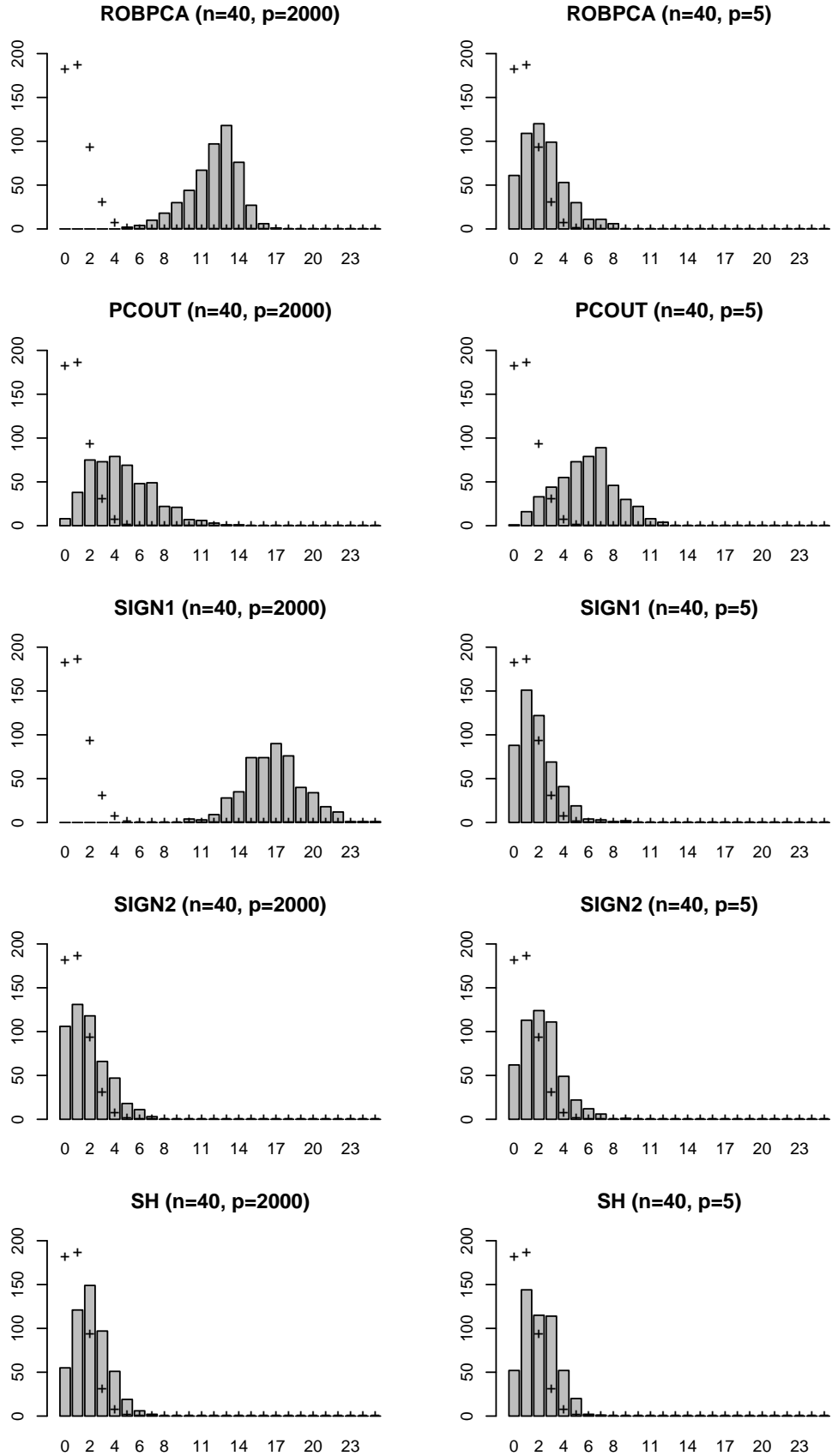}
\end{center}
\caption{Results of the detection of outliers by various methods (ROBPCA, PCOUT, SIGN1, SIGN2 and SH, see text for details) in 500 simulated multivariate normal data sets with n = 40 observations and p = 2000 variables (left) and p = 5 variables (right). The '+' signs represent expected frequencies under a binomial(40,0.025) distribution.}
\label{fig4}
\end{figure}

We conclude that, except for the PCOUT method, all the observed frequencies have moved closer to the expected frequencies and in the case of ROBPCA and SIGN1 there is in fact a fantastic recovery. In this low dimensional situation all the five methods but one are in agreement and doing more or less close to what they are supposed to do.

As we wondered what could possibly justify the observed behaviour of the five methods
we looked into the algorithmic details of each method and can then add the following explanation for the three types of behaviour observed.

Methods that work fine in low dimensions and fail in high dimensions (ROBPCA and SIGN1): when $p\ge n$ both methods work internally with the data rotated to its proper subspace (of dimension $n-1$) and then look for projections showing outliers. But, an easy application of Theorem 2 to one-dimensional projections shows that (as long as $p\ge n-1$) there is at least one projection where each and every observation sticks out from the remaining, which are projected all to the same point. This makes the usual criterion based on robust z-scores (standardized observations using the median and the median absolute deviation in place of the mean and the standard deviation) or the more sophisticated Stahel--Donoho outlyingness measure (for each observation this  is the maximum of all possible robust z-scores of that observation, taken along all possible projections) completely useless. When the number of variables is small ($p\ll n$) there are no longer such ``arbitrary'' projections and both methods work.

Methods that work fine in low and high dimensions (SIGN2 and SH): SIGN2 is similar to SIGN1, and SH is similar to ROBPCA. However, in both cases a dimension reduction by principal components is the first step of the procedures. This is then followed by  outlier detection in the reduced space (the number of  principal components to retain is chosen by fixing a proportion of explained variance, in the case of SIGN1, or by an automatic selection method based on the scree plot, in the case of SH). By bringing the high dimensional situation to the usual $p\ll n$ the difficulties described above are avoided, and therefore the methods work as expected no matter the dimensionality.

Method with problems both in high and low dimensions (PCOUT): according to \citet{Filzmoser:Maronna:Werner:2008}, PCOUT was designed to work  in high dimension cases. This explains why it does not work very well in low dimensions. However, in the high dimensional situation, the method uses robust versions of the Mahalanobis distance, which, in the light of Theorem 1
  is very dangerous, and may explain the problems encountered.

This example clearly shows that knowledge of the geometric properties of high dimensional data, in particular the results given in this paper,  is of crucial importance for the correct development of methods to deal with this type of data.

\section{Conclusions} 
New technologies are great in providing floods of data full of information and potential knowledge, but the extraction of such information can often prove very difficult. High dimensional spaces are typical in revealing such difficulties that mathematicians label curse of dimensionality. In the present work we focus on  spaces where the number of variables ($p$) is at least the number of observations ($n$), a case that occurs, in statistical  practice, mostly within the high dimensional data framework. We look at the behaviour of the Mahalanobis distance and the idea of projection, so essentials in the analysis of multivariate data.

When $p\ge n-1$ we prove that, the Mahalanobis distance becomes degenerated and  its known distance properties are lost. Under the same restriction it is confirmed that the idea of projection loses interest and becomes mostly useless.  These conclusions imply that many procedures for analyzing data will not work in these spaces.

Although we have seen users insisting in using artifacts to analyze this kind of data, it seems that they are not aware of the geometric complexities of spaces under these conditions. Disclosing what happens to the Mahalanobis distance and to the usual system of projections we gain insight into the geometric properties of those spaces and hopefully contribute to: (i) prevent unconscious applications of inadequate methods and (ii) help to devise new methods where those geometric properties have to be taken into account.

Having said this, it should be understood that we are not suggesting that traditional statistical methods are not worthwhile and their use should be abandoned when analyzing high dimensional data. On the contrary, we consider that this formidable source of statistical knowledge should not be forgotten but used properly. To allow traditional methods to operate one first step towards reducing the dimensionality of the data should be given. For dimensionality reduction to be achieved one can think of a number of techniques including a panoply of variable selection  and regularization procedures.

\section*{Acknowledgement}
 This work received financial support from Portuguese National Funds through Funda\c{c}\~{a}o para a Ci\^{e}ncia e a Tecnologia. 


\appendix

\section*{Appendix}
\subsection*{Definitions and notation}

\noindent
\begin{definition}[notation] Data matrix ($n\times p$): $\displaystyle X= \left( 
\begin{array}{c}
	x_1^T \\ \cdots \\ x_n^T
\end{array}
 \right) = \left( x_1 \cdots x_n \right)^T$. 
 \end{definition}

\begin{definition}[points in general position]

 The p-dimensional observations in the ($n\times p$) data matrix $X$ are said to be in general position if $X$ has maximal rank, that is, if ${\rm rank}(X)=\min(p,n)$. This property holds with probability one for continuous variables.  

\end{definition}

\begin{definition}[further notation]
Mean of $X$ ($n \times 1$): $ \overline{x}= \sum_{i=1}^n x_i/n$.
 
Centered data matrix ($n \times p$): $X_c= \left( x_1-\overline{x} \cdots x_n-\overline{x} \right)^T=\left( I_n -v_1 v_1^T/n \right)X$ ($v_x$ denotes a vector with all its elements equal to $x$). 

Covariance matrix of $X$  ($p \times p$): $ S=  \sum_{i=1}^n (x_i-\overline{x})(x_i -\overline{x})^T/(n-1) = X_c^TX_c /(n-1)$. If $X$ is in general position then rank$(S)=$rank$(X_c)=\min(p,n-1)$.

Standardized data matrix ($n \times p$): $Z = X_c S^{-1/2}$, where $S^{-1/2}$ is any square root of $S^{-1}$ (which means that $S^{-1/2}S^{-1/2}=S^{-1}$).

Mahalanobis distance between an observation, ${x}_i$, and $\overline{x}$: $d_{{x}_i,\bar{x}}= \left\{ ({x}_i - \bar{x})^T S^{-1}({x}_i - \bar{x})  \right\}^{1/2}$. 

Mahalanobis distance between two observations, ${x}_i$ and ${x}_j$: $d_{{x}_i,{x}_j}=\left\{ ({x}_i - {x}_j)^T S^{-1}({x}_i - {x}_j) \right\}^{1/2}$. 

Matrix $D$ ($n \times n$): 
\begin{equation}
		D = X_c S^{-1} X_c^T =(n-1) X_c (X_c^T X_c)^{-1} X_c^T = Z Z^T,
\end{equation}
	where $d_{ij}= (x_i - \overline{x})^T  S^{-1}(x_j - \overline{x})$. Note that 
	$d^2_{x_i,\overline{x}} = d_{ii}$ and  $d^2_{x_i,x_j}= d_{ii}+ d_{jj}-2 \, d_{ij}$.

Augmented data matrix ($p \times (n+1)$): 
	$X_d = \left[  \begin{array}{c|c} v_1 & X_c  \end{array}  \right]$, which verifies
\begin{equation}
\label{relHD}
	\displaystyle X_d (X_d^T X_d)^{-1} X_d^T = H = \frac{D}{n-1}+ \frac{v_1v_1^T}{n},
\end{equation}
where $H$ is the well known ``hat matrix'' \citep{Hoaglin:Welsch:1978}.
\end{definition}

\subsection*{Main results}

\setcounter{theorem}{0}

\begin{theorem} 
For every dataset with $n$ points,  $x_1, \ldots, x_n$ (in general position), and $p \ge n-1$  variables we have that: (i)~$d_{{x}_i,\bar{x}}=(n-1) \; {n}^{-1/2}$, for every $i=1, \ldots , n$, and (ii)~$d_{{x}_i,{x}_j}= \left\{ 2(n-1) \right\}^{1/2}$ for every $i\ne j=1, \ldots , n$.
\end{theorem}

\textbf{Proof:}
For $p > n-1$, $S$ is singular and must be replaced by a generalized inverse ($S^-$). This is equivalent to reducing the number of variables to $n-1$  (\textit{e.g.}, the standardized non-trivial principal components). Therefore we only have to prove  the $p=n-1$ case.

In that case, $X_c$ has dimension $n\times (n-1)$ and rank $n-1$, whereas $X_d$ has dimension $n\times n$ and rank $n$, so it can be inverted, resulting in
\[
 H = X_d (X_d^T X_d)^{-1} X_d^T = X_d X_d^{-1}(X_d^T)^{-1} X_d^T =I = \frac{D}{n-1}+ \frac{v_1v_1^T}{n},
\]
which is equivalent to
\begin{equation} \label{resD}
	 D = (n-1)\left( I - \frac{v_1v_1^T}{n} \right) .
\end{equation}
That is, $\displaystyle d_{ii}=(n-1)\left( 1 - 1/n \right)= (n-1)^2/n $, for all $i$, which proves (i).
On the other hand, from (\ref{resD}), for any  $i \ne j$, $\displaystyle d_{ij}= - (n-1)/n$, therefore
\[
\displaystyle d^2_{x_i,x_j}= d_{ii}+ d_{jj}-2 \, d_{ij} = 2\, \frac{(n-1)^2}{n} + 2\,\frac{n-1}{n} = 2 \, (n-1),
\]
which proves (ii).

\setcounter{theorem}{0}

\begin{theorem}[extra results] For every dataset with $n$ points,  $x_1, \ldots, x_n$,  and $p$  variables we have that: (iii)~$d_{{x}_i,\bar{x}}\le (n-1)\; n^{-1/2}$, for every $i=1, \ldots , n$, and (iv)~$d_{{x}_i,{x}_j}\le  \left\{ 2(n-1) \right\}^{1/2}$ for every $i\ne j=1, \ldots , n$.
\end{theorem}

\textbf{Proof:} 
 (iii) Equation (\ref{relHD}) writes $h_{ij}=d_{ij}/(n-1)+1/n$, and applying the well known result \citep{Hoaglin:Welsch:1978}, $0\le h_{ii} \le 1$, the conclusion that $d_{ii} \le (n-1)^2/n$ follows immediately \citep[this is a short alternative proof of Theorem 2.1 in][]{Gath:Hayes:2006}.

\noindent
(iv) Define $T= I - H$ and check that $T$ is symmetric and idempotent. Consider the $i$th and $j$th row (or column) vectors of $T$, $t_i$ and $t_j$, with $i \ne j$. Then $t_i^T t_j=-h_{ij}$ and $t_i^T t_i=1-h_{ii}$ (we  need the assumption that $h_{ii}<1$, for all $i$, which is not restrictive, because the cases where $h_{ii}=1$ are considered in (ii)). Let $\beta$ be the angle between $t_i$ and $t_j$, then
\[
\cos \beta = \frac{t_i^T t_j}{\left( t_i^T t_i \right)^{1/2} \left(t_j^T t_j\right)^{1/2}}= - \frac{h_{ij}}{\left(1-h_{ii}\right)^{1/2}\left(1-h_{jj}\right)^{1/2}},
\]
implying that $- \left( 1-h_{ii} \right)^{1/2} \left( 1-h_{jj} \right)^{1/2} \le h_{ij} \le \left( 1-h_{ii} \right)^{1/2} \left( 1-h_{jj} \right)^{1/2} $. Considering this result and the  inequality $\left\{ \left( 1-h_{ii} \right)^{1/2} - \left( 1-h_{jj} \right)^{1/2} \right\}^2 \ge 0$, it follows that
\[
0 \le 1-h_{ii} + 1-h_{jj} - 2 \left( 1-h_{ii} \right)^{1/2} \left( 1-h_{jj} \right)^{1/2}  \le 1-h_{ii} + 1-h_{jj} +2\, h_{ij},
\]
which is equivalent to $h_{ii} + h_{jj} - 2 \, h_{ij} \le 2$.  Finally, simple manipulations of the definitions yield
\[
d^2_{x_i,x_j}= d_{ii}+ d_{jj}-2 \, d_{ij} = (n-1) (h_{ii}+ h_{jj}-2 \, h_{ij}) \le 2 (n-1) .
\]

\begin{theorem} For every dataset with $n$ points and $p \ge n-1$ variables, $X$, and every non-singular two dimensional arrangement of $n$ points ($Y$) it is always possible to find (explicitly) an orthogonal projection of  $X$ which is ``similar'' to $Y$. More precisely, given $X$ and $Y$, there is an orthogonal transformation, characterized by the $p\times 2$ matrix $Q$, whose columns are orthonormal vectors, such that $X Q =Y^\ast$, where $Y^\ast$ is an affine transformation of $Y$ (that is, $Y^\ast = Y A + v_1b^T$, for some non-singular $2\times 2$ matrix $A$ and some two dimensional vector $b$).
\end{theorem}

\textbf{Proof:}
Without loss of generality we assume that $Y$ $(n \times 2)$ is such that $\overline{y}=v_0$ and 
$S_{Y} = Y^T Y/(n-1) = I_2$. As in the proof of Theorem 1 we need to consider only the case $p =n-1$, for which the covariance matrix of $X$, $S$, is invertible. Let $Z = X_c S^{-1/2}$, be the standardized data matrix, and consider the  $(n-1)\times 2$ matrix $U$ defined by $U = Z^T Y/(n-1)$. We show next that (i) the columns of $U$ are orthonormal vectors, and (ii) $Y = Z U$.

\noindent
(i) $U^T U = Y^T Z Z^T Y/(n-1)^2 = Y^T D Y/(n-1)^2 = Y^T \left( I_n - v_1v_1^T/n  \right) Y/(n-1)=Y_c^T Y_c/(n-1) =I_2 $, which follows from equation (\ref{resD}) and by the assumption that $Y$ is standardized.

\noindent
(ii) $Z U=Z Z^T Y/(n-1) = D Y/(n-1) = \left( I_n - v_1v_1^T/n  \right) Y = Y_c =Y$, for the same reasons.

\noindent
The next step is to rewrite (ii) in terms of $X$ or, equivalently, in terms of $X_c$,
$Y = Z U = X_c S^{-1/2} U$. Replacing, in the last equality, the matrix $S^{-1/2} U$ by its singular value decomposition, $S^{-1/2} U= V_1 L V_2$,  where the columns of $V_1$ ($p\times 2$) and $V_2$ ($2 \times 2$) are orthonormal and  $L$ is a $2\times 2$ diagonal matrix, leads to $Y = X_c V_1 L V_2$, which is equivalent to
$Y V_2^T L^{-1} = X_c V_1 $. Therefore, and concluding the proof, $Q= V_1$, $A=V_2^T L^{-1}$ and $b=V_1^T \overline{x}$.

\begin{remark}
The two-dimensional projections are usually the most interesting to consider, but the theorem could be generalized easily to projections on a subspace of arbitrary dimension, $k=1, \ldots , n-2$. The case $k=1$ would contain the ``piling effect'' as a special case, and also the curious case of projections where all but one point (which can be any of the $n$ points) coincide.
\end{remark}

\begin{remark}
The key to  prove both theorems is the fact that $H=I_n$ for $p \ge n-1$.
\end{remark}
	
\begin{remark}
The plots in Figure~2, produced with the formulas given in this theorem, illustrate the kind of ``similarity'' between $Y$ and $Y^\ast$ that can be obtained when $p \ge n-1$, no matter the data, and that can be described as perfect ``similarity''.  On the contrary, in the cases considered in Figure~3 where $p < n-1$, perfect ``similarity'' is no longer possible. To produce plots as ``similar'' as possible to $Y$ we used a least squares criterion and look for the transformation, $X Q=\hat{Y}^\ast$, minimizing $\| Y^\ast - \hat{Y}^\ast \|$, concluding that  $U$ must be replaced by
$U = \left(Z^T Z\right)^{-1} Z^T Y/(n-1)$ (all the other formulas are unchanged).
\end{remark}

\vspace*{1.4cm}
\subsection*{Supplementary material: Example materials and methods}

The data used in this example, described in \citet{Alonal:1999}, were downloaded from the  Princeton University Gene Expression Project database \\
 \texttt{\footnotesize http://genomics-pubs.princeton.edu/oncology/ affydata/index.html}. \\
 The data were then imported into \texttt{R} \citep{R:2010} where all the computations were performed. The only preprocessing applied was a $\log_2$ transform of the full data matrix \citep[following][]{Shieh:Hung:2009}. 
 
 The data matrix has 62 rows (samples or cases) and 2000 columns (genes). The 62 cases are divided into two groups, the normal tissue samples (22, labeled N\ldots) and the tumor samples (40, labeled T\ldots). 
 
The purpose of the example is to assess the performance of some outlier detecting methods recently proposed and advertised as appropriate for high dimensional datasets. The  
methods considered are the ROBPCA of \citet{Hubert:Rousseeuw:VandenBranden:2005}, three methods proposed in \citet{Filzmoser:Maronna:Werner:2008}, denominated by the authors as PCOUT, SIGN1 and SIGN2, and the method proposed in \citet{Shieh:Hung:2009}, referred to as SH. 

In the application of these methods we have used the following implementations: function \texttt{PcaHubert}, from package \texttt{rrcov}, with default settings, for ROBPCA; functions \texttt{pcout}, \texttt{sign1} and \texttt{sign2} from package \texttt{mvoutlier}, all with their default settings; and the script written and made available by the authors for SH \citep[again with default settings, see][]{Shieh:Hung:2009}.

Each method was applied  to each group separately, and  the outliers detected were in number as referred to in the main text. To complement that information we provide here the labels of those outlying observations:

\begin{description}
\footnotesize
\item[\rm ROBPCA:] N8 N9 N12 N34 N36 / T2 T4 T5 T6 T9 T12 T19 T25 T34 T37;
\item[\rm PCOUT:] N12 / T5 T30 T33 T36 T37 T39; 
\item[\rm SIGN1:] N3 N8 N9 N10 N12 N29 N34 /  T2  T5  T6  T9 T10 T12 T17 T18 T19 T20 T21 T25 T26 T28 T29 T30 T31 T32 T34 T37 T38; 
\item[\rm SIGN2:] N8 N9 N12 N34 /  T2  T5  T6  T9 T12 T29 T32 T37; 
\item[\rm SH:] N8 N12 N34 / T2 T30 T33 T36 T37.
\end{description}

\noindent
The Monte Carlo simulation described in the main text was performed with the following scripts:

{ \scriptsize
\begin{verbatim}
###
### This script simulates the detection of outliers using the ROBPCA method of Hubert et al (2009), 
### It requires library rrcov
### t(colon.orig) contains the colon data matrix  (after the log(,2) transformation
### new.y contains indication of the groups (1 for normal and 2 for tumor)
### The results are stored in the vectors out22.h and out40.h
###

require(rrcov)

### group principal components and means of the original dataset 
### (to be replicated in the simulated data)
 
pca.cdI<-prcomp(t(colon.orig)[new.y==1,])
pca.cdII<-prcomp(t(colon.orig)[new.y==2,])
mean.1<-colMeans(t(colon.orig[,new.y==1]))
mean.2<-colMeans(t(colon.orig[,new.y==2]))

set.seed(200)
outs22.h<-rep(0,500)
outs40.h<-rep(0,500)
for (i in 1:500){
auxX1<-mvrnorm(22,mu=rep(0,21),Sigma=diag(21),empirical=T)%*%diag(pca.cdI$sdev[-22])
                                                                 %*%t(pca.cdI$rot[,-22])
auxX2<-mvrnorm(40,mu=rep(0,39),Sigma=diag(39),empirical=T)%*%diag(pca.cdII$sdev[-40])
                                                                 %*%t(pca.cdII$rot[,-40])
auxX1<-t(t(auxX1)+mean.1)
auxX2<-t(t(auxX2)+mean.2)
pca.auxX1<-PcaHubert(auxX1)
pca.auxX2<-PcaHubert(auxX2)
outs22.h[i]<-sum(pca.auxX1@flag==F)
outs40.h[i]<-sum(pca.auxX2@flag==F)
}

###
### This script simulates the detection of outliers using the methods of Filzmoser et al (2008)
### It requires library mvoutlier
### The results are stored in the vectors outs.cpII.22.f1, outs.cpII.22.f2, outs.cpII.22.f3, 
### outs.cpII.40.f1, outs.cpII.40.f2, and outs.cpII.40.f3
###

set.seed(200)
outs.cpII.22.f1<-rep(0,500)
outs.cpII.22.f2<-rep(0,500)
outs.cpII.22.f3<-rep(0,500)
outs.cpII.40.f1<-rep(0,500)
outs.cpII.40.f2<-rep(0,500)
outs.cpII.40.f3<-rep(0,500)
for (i in 1:500){
auxX1<-mvrnorm(22,mu=rep(0,21),Sigma=diag(21),empirical=T)%*%diag(pca.cdI$sdev[-22])
                                                                 %*%t(pca.cdI$rot[,-22])
auxX2<-mvrnorm(40,mu=rep(0,39),Sigma=diag(39),empirical=T)%*%diag(pca.cdII$sdev[-40])
                                                                 %*%t(pca.cdII$rot[,-40])
auxX1<-t(t(auxX1)+mean.1)
auxX2<-t(t(auxX2)+mean.2)
outs.cpII.22.f1[i]<-sum(pcout(auxX1)$wfinal01==0)
outs.cpII.40.f1[i]<-sum(pcout(auxX2)$wfinal01==0)
outs.cpII.22.f2[i]<-sum(sign1(auxX1)$wfinal01==0)
outs.cpII.40.f2[i]<-sum(sign1(auxX2)$wfinal01==0)
outs.cpII.22.f3[i]<-sum(sign2(auxX1)$wfinal01==0)
outs.cpII.40.f3[i]<-sum(sign2(auxX2)$wfinal01==0)
}

###
### This script simulates the detection of outliers using the method of Shieh snd Hung (2009)
### It requires the script of the authors (outlier) which in turn requires library rrcov
### The results are stored in the vectors out22.sh and out40.sh
###

set.seed(200)
outs22.sh<-rep(0,500)
outs40.sh<-rep(0,500)
grupos<-c(rep(1,22),rep(2,40))
for (i in 1:500){
auxX1<-mvrnorm(22,mu=rep(0,21),Sigma=diag(21),empirical=T)%*%diag(pca.cdI$sdev[-22])
                                                                 %*%t(pca.cdI$rot[,-22])
auxX2<-mvrnorm(40,mu=rep(0,39),Sigma=diag(39),empirical=T)%*%diag(pca.cdII$sdev[-40])
                                                                 %*%t(pca.cdII$rot[,-40])
auxX1<-t(t(auxX1)+mean.1)
auxX2<-t(t(auxX2)+mean.2)
auxX<-rbind(auxX1,auxX2)
res.out<-outlier(auxX,grupos)
outs22.sh[i]<-sum(grupos[res.out]==1)
outs40.sh[i]<-sum(grupos[res.out]==2)
}
\end{verbatim}
}

\noindent
The results when $p=2000$ (summarized with \texttt{table()}) were the following:

{ \scriptsize
\begin{verbatim}
outs40.h
  5   6   7   8   9  10  11  12  13  14  15  16  17 
  2   4  10  18  30  44  67  97 118  76  27   6   1 
outs22.h
  2   3   4   5   6   7   8 
  2  11  75 127 213  68   4 


outs.cpII.40.f1
 0  1  2  3  4  5  6  7  8  9 10 11 12 13 14 
 8 38 75 73 79 69 48 49 22 21  7  6  3  1  1 
outs.cpII.22.f1
  0   1   2   3   4   5   6   7   8   9 
 27  83  92 101  85  51  34  18   8   1 

outs.cpII.40.f2
10 11 12 13 14 15 16 17 18 19 20 21 22 23 24 25 
 4  3  9 28 35 74 74 90 76 40 34 18 12  1  1  1 
outs.cpII.22.f2
 2  3  4  5  6  7  8  9 10 11 12 13 14 15 17 
 1  6 17 42 58 99 94 79 50 27 14  6  4  2  1 

outs.cpII.40.f3
  0   1   2   3   4   5   6   7 
106 131 118  66  47  18  11   3 
outs.cpII.22.f3
  0   1   2   3   4   5 
115 142 123  76  32  12 

outs40.sh
  0   1   2   3   4   5   6   7 
 55 121 149  97  51  19   6   2 
outs22.sh
  0   1   2   3   4   5   6 
 63 133 142  94  53  13   2 
 \end{verbatim}
}

For $p=5$, the 5 variables were selected randomly (with a call to function sample) and the previous code was run after changing only the lines applying the methods (for instance, \texttt{pca.auxX1$<-$PcaHubert(auxX1[,xvar])}, where \texttt{xvar} contains the labels of the selected five variables). The results  were the following:

{ \scriptsize
\begin{verbatim}
outs40.h
  0   1   2   3   4   5   6   7   8 
 61 109 120  99  53  30  11  11   6 
outs22.h
  0   1   2   3   4   5   6   7   8 
 34  62 120 152  77  41  10   3   1 

outs.cpII.40.f1
 0  1  2  3  4  5  6  7  8  9 10 11 12 
 1 16 33 44 55 73 79 89 46 30 22  8  4 
outs.cpII.22.f1
  0   1   2   3   4   5   6   7 
 11  52  67 116 106  91  41  16 

outs.cpII.40.f2
  0   1   2   3   4   5   6   7   8   9 
 88 151 122  69  41  19   4   3   1   2 
outs.cpII.22.f2
  0   1   2   3   4   5   6   7   9 
145 143 109  48  29  18   6   1   1 

outs.cpII.40.f3
  0   1   2   3   4   5   6   7   9 
 62 113 124 111  49  22  12   6   1 
outs.cpII.22.f3
  0   1   2   3   4   5   6   7 
114 134 136  70  36   6   3   1 

outs40.sh
  0   1   2   3   4   5   6   7 
 52 144 115 114  52  20   2   1 
outs22.sh
  0   1   2   3   4   5   6   7 
 46 131 135 100  61  22   4   1 
 \end{verbatim}
}
 
 \end{document}